\documentclass[preprint]{aastex}

\newcommand\degd{\ifmmode^{\circ}\!\!\!.\,\else$^{\circ}\!\!\!.\,$\fi}

\newcommand{\lsim}{\stackrel{\scriptstyle <}{\scriptstyle \sim}}
\newcommand{\gsim}{\stackrel{\scriptstyle >}{\scriptstyle \sim}}

\shortauthors{Bower et al.}
\shorttitle{RIPL II:  GJ 896A}

\begin{document}

\title{Radio Interferometric Planet Search II:  Constraints on sub-Jupiter-Mass Companions to GJ 896A}
\author{Geoffrey C. Bower\altaffilmark{1}, Alberto Bolatto\altaffilmark{1,2}, Eric B. Ford\altaffilmark{1,3}, Adam Fries\altaffilmark{1}, Paul Kalas\altaffilmark{1}, Karol Sanchez\altaffilmark{1}, Phoebe Sanderbeck\altaffilmark{4}, Vincent Viscomi\altaffilmark{1}}
\email{gbower@astro.berkeley.edu}

\altaffiltext{1}{Astronomy Department \& Radio Astronomy Laboratory,University of California, Berkeley, CA 94720}
\altaffiltext{2}{Department of Astronomy, University of Maryland, College Park, MD, 20742-2421}
\altaffiltext{3}{Astronomy Department, University of Florida, Gainesville, FL 32611-2055}
\altaffiltext{4}{Department of Physics, Brandeis University, Abelson-Bass-Yalem 107, MS 057,  415 South Street, Waltham, MA 02453 }

\begin{abstract}
We present results from the Radio Interferometric Planet (RIPL) search for companions to the nearby star GJ 896A.  
We present 11 observations over 4.9 years.  Fitting astrometric parameters to the data reveals
a residual with peak-to-peak amplitude of $\sim 3$ mas in right ascension.  This residual is well-fit by an acceleration
term of $0.458 \pm 0.032$ mas/y$^2$.  The parallax is fit to an accuracy of 0.2 mas and the proper motion
terms are fit to accuracies of 0.01 mas/y.  After fitting astrometric and acceleration
terms residuals are 0.26 mas in each coordinate, demonstrating that stellar jitter does not limit the ability to
carry out radio astrometric planet detection and characterization.  
The acceleration term originates in part from the companion
GJ 896B but the amplitude of the acceleration in declination is not accurately predicted by the orbital model.
The acceleration sets a mass upper limit of 0.15 $M_J$ at a semi-major axis of 2 AU 
for a planetary companion to GJ 896A.  For semi-major axes between 
0.3 and 2 AU upper limits are determined by the maximum angular separation; the upper limits 
scale from the minimum value in proportion to the inverse of the radius.  Upper limits at larger radii are 
set by the acceleration and scale as the radius squared.
An improved solution for the stellar binary system could improve 
the exoplanet mass sensitivity by an order of magnitude.
\end{abstract}


\section{Introduction}

The study of extrasolar planets provides an important link between the
study of star formation, the study of our own solar system, and the
search for extraterrestrial life.  Radial velocity searches \citep{2006ApJ...646..505B},
transit searches with Kepler \citep{2011ApJ...728..117B}, and microlensing searches \citep{2008Sci...319..927G} 
have surveyed significant components of the exoplanet mass-separation parameter space
exposing a wide variety of phenomenology.

Despite their abundance,
low mass stars have been less well-characterized for a number of reasons.
In particular, low mass stars are often too faint or too strongly
variable for high precision radial velocity and transit searches.
Microlensing searches have identified companions to low mass
stars \citep{2008ApJ...684..663B,2006Natur.439..437B,2006ApJ...644L..37G}, 
which suggest that such companions may be very common.  On the other
hand, radial velocity searches suggest very low planet occurrence rates
for close-in companions
\citep{2006ApJ...649..436E,2007ApJ...670..833J}.  The frequency and
distribution of companions to low mass stars can be an important diagnostic
of methods for planet formation
\citep{2004ApJ...612L..73L,2005ApJ...626.1045I,2006ApJ...643..501B,2007Ap&SS.311....9K}.

Astrometry can be a powerful tool for detection and characterization of 
nearby, low-mass planetary systems.
\citet{2010AJ....140.1657M} demonstrate
that optical astrometric measurements with the Palomar Testbed Interferometer
are capable of achieving sub-milliarcsecond accuracy in the separation between
close binaries.  They summarize
the known properties of sub-stellar companions to low mass stars in binary systems and
conclude that giant planet companions in such systems are common.  
The Carnegie Astrometric Planet Search Program \citep{2009PASP..121.1218B}
has also obtained sub-milliarcsecond astrometry on the low mass star 
NLTT 48256 over 2 years of observations;
this program will ultimately survey 100 nearby M, L, and T dwarfs.

The Very Long Baseline Array
(VLBA) can routinely achieve an astrometric accuracy of $\sim100$
$\mu$as in a single epoch at frequencies $\lsim 10$ GHz and $\sim 20$ $\mu$as at higher frequencies. 
The VLBA is capable of accuracies as high as 8 $\mu$as under
favorable circumstances \citep{2003ApJ...598..704F}.   To use this
technique the target source must have a sufficiently high brightness
temperature to be detected by a high resolution radio
interferometer.  For the VLBA, brightness temperatures must be
$T_b \gsim 10^{7}$ K, requiring nonthermal emission.  Thus, the
systems which can be studied are limited to the most active stars.
VLBA astrometry has been important for a wide range of 
stellar measurements including parallax \citep[e.g.,][]{2007ApJ...667.1161S,2007A&A...474..515M,2007ApJ...671..546L,2009ApJ...698..242T},
galactic structure  \citep{2006Sci...311...54X,2011ApJ...733...25X,2011AN....332..461B},
binary orbit characterization
\citep{2006A&A...446..733G}, and the radio/optical reference-frame link
\citep{1999A&A...344.1014L}. 
Recently, \citet{2009ApJ...706L.205F} detected a brown dwarf with the 
VLBA and explored the possibilities of companion detection.
\citet{2007arXiv0704.0238B} reviewed the prospects for future radio astrometric surveys with
respect to other methods.

RIPL, the Radio Interferometric Planet Search, is a 4-year program 
using the VLBA and the 100m Green Bank Telescope
(GBT) to search for the astrometic signatures of massive planets 
around nearby ($D < 10$ pc), low-mass (M dwarf) stars.  
We are observing a sample of 30 stars a total of
12 times over a 4-year period with the expectation of achieving $\sim 0.1$ milliarcsecond
astrometric accuracy per epoch, and ultimately having the sensitivity to 
detect Jupiter-mass planets at 1 AU.  This astrometric search is an important 
complement to radial velocity, transit, microlensing, and optical astrometric searches for planets on 
account of greater sensitivity to  1)
low mass stars, 2) long-period
planets, 3) active stars, and, 4) 
planets that may be imaged with extreme adaptive optics.  
Many RIPL stars are
in binary systems, which may have higher planetary fractions.

In Paper I \citep{2009ApJ...701.1922B}, we used Very Large Array (VLA) observations to identify
the sample of stars appropriate for RIPL and we used the VLBA
to characterize the detectability and short-term ($\lsim 10$ days) stability of 
the astrometric images of these stars.  The short-term stability addresses a significant
concern of radio astrometry that stellar activity will lead to large astrometric
errors.  The absence of a significant difference
in the historical optical proper motions and the radio proper motion from
this RIPL precursor survey permitted us to exclude brown dwarf
companions to the four stars observed.  RIPL observing began in October 2007
and is anticipated to end in late 2011.

In this paper, we present initial results from RIPL for the star GJ 896A.
This star was also observed as part of the precursor survey in paper I.
Together these observations demonstrate astrometric stability over a 4.9 year
span.
GJ 896A (EQ Peg A) is a variable star in a multiple system with GJ 896 B, C, and D.  
GJ 896B is its closest companion; the orbit has been characterized, but it is not well
constrained 
\citep{1984AJ.....89.1063H}.
GJ 896 A and B have spectral types M3.5 and M4.5.  These two stars have been detected
previously in the radio and seen to have point-like structure and strong radio variability
\citep{1985A&A...149...95P,1989A&A...210..284J,1995A&A...298..187B,1998ASPC..154.1484G}.  
In \S 2, we describe our observations, analysis, and results.  In \S 3, 
we present our astrometric analysis of the data.  In \S 4, we examine the constraints that
we can place on companions to GJ 896A.  In \S 5, we summarize our results.

\section{Observations, Data Analysis, and Results}

Observations of GJ 896A were obtained as part of the RIPL survey.  RIPL observations spanned 
October 2007 through June 2011, and are ongoing.  In each epoch, we observed two stars over
a span of 8 hours.  Fast switching with a phase reference calibrator and secondary
(and sometimes tertiary) astrometric calibrators reduced total observing
time on each star to approximately two hours.  The total span from start to end of
observation for a particular star within an epoch varied, depending on the relative
coordinates of the stars and the allocated time slot.  In some instances, the stars were observed sequentially
in four hour blocks.  In other instances, the stars were interleaved and the span was over
the full eight hours.

Observations typically included all ten antennas of the Very Long Baseline Array as
well as the 100 meter Green Bank telescope.  Data were recorded at a rate of 512 Mbps
in dual circular polarization mode with observing bands at 8.4 GHz using the standard VLBA mode
v4cm-512-8-2.  More precisely, this mode corresponds to four dual-polarization frequency bands centered 
at 8384.49, 8400.49, 8416.49, and 8432.49 MHz, each sampled at 32 Msamples/second with 2 bit sampling.
Occasional observations of
fringe calibrators were interspersed throughout each epoch.  The switching cycle 
included repeated observations of 1 minute on the phase reference calibrator followed by 
3 minutes on the star.  Every 30 minutes short duration observations of the secondary
and tertiary calibrators phase referenced to the primary calibrator were obtained.

Data were reduced with an AIPS \citep{2003ASSL..285..109G}
pipeline that performed standard calibrations of
amplitude and phase.  Corrections for the atmospheric total electron content (TEC)
and post-correlation updates to the Earth orientation parameters (EOP) were also
included.  All sources were imaged and deconcolved using the CLEAN algorithm.
The typical synthesized beam for GJ 896A was 3.0 $\times$ 1.1 mas, oriented North-South.
We fit Gaussian models to all sources using the AIPS task JMFIT.

\begin{figure}
\includegraphics[width=1.0\textwidth]{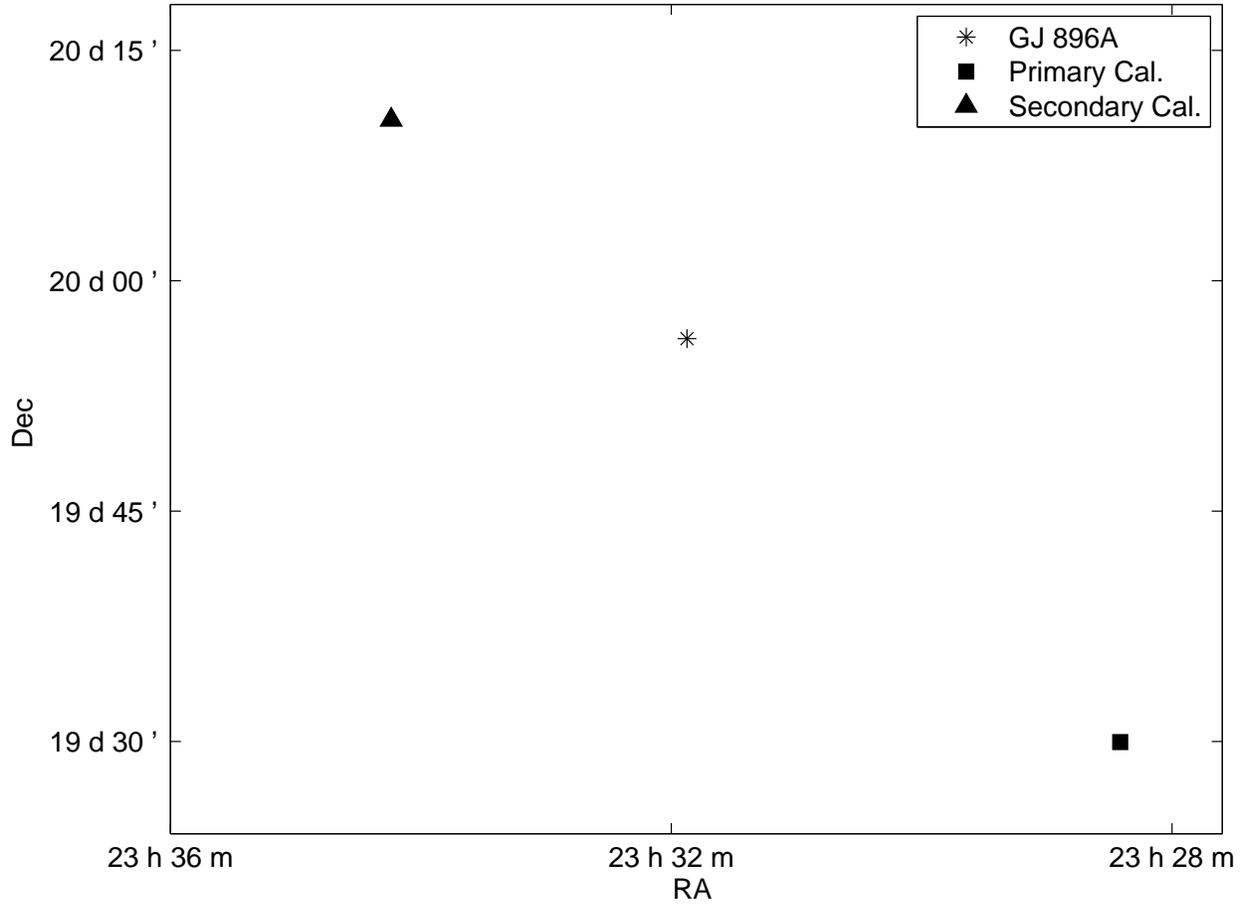}
\caption{Positions for GJ 896A and primary and secondary calibrators.
The distance from the primary calibrator, J2328+1929, to GJ 896A is 0.9 deg.
The distance from the secondary calibrator, J2334+2010, to GJ 896A is 0.6 deg.
The nearly linear arrangement and small separations make this an
excellent set of calibrators for this star.
\label{fig:pos}
}
\end{figure}

In the case of GJ 896A, the primary phase reference calibrator is J2328+1929, which
has an ICRF2 position that is accurate to 1.84 mas $\times$ 1.58 mas 
\footnote{\tt http://gemini.gsfc.nasa.gov/solutions/2010a/2010a.html}.  
The assumed position for J2328+1929 is (23$^h$ 28$^m$ 24.874773$^s$,  19$^\circ$ 29\arcmin 58.03010\arcsec);
these updated values are slightly offset from the assumed position of paper I by (0.254, 0.059) mas 
in right ascension and declination, respectively.  All positions for other sources are relative to
this position.
The secondary calibrator is J2334+2010.  No tertiary calibrator was used.  
We apply a correction for tropospheric phase gradients on the
astrometric position of the star by removing variations in the position of 
the secondary calibrator.  This correction is applied on an epoch-by-epoch basis.  
A more sophisticated approach using the AIPS task ATMCA can make corrections
on shorter timescales.  The magnitude of corrections to GJ 896A are 
$\sim 0.1$ mas.  This approach may overestimate trophospheric errors slightly given
the greater distance and higher declination of J2334+2010 from the primary
calibrator.  The close spacing and nearly linear arrangement of the star and
calibrators, however, make this a good calibration set (Fig.~\ref{fig:pos}).
Typically,
3C 454.3 was observed as the fringe calibrator.  As of June 2011, we have 8 good epochs
for GJ 896A.

Both calibrators are dominated by a strong point source and show only weak 
evidence for non-point-like structure.  Images of the calibrators are
included in Paper I.  The primary calibrator, J2328+1929,
has a typical flux density of 60 mJy; the secondary calibrator, J2334+2010,
has a typical flux density of 20 mJy.  In Table~\ref{tab:secondary}, 
we summarize flux densities and positions for the secondary calibrator as 
a function of epoch; we also include the mean value from the three epochs of paper I.
The positions for J2334+2010 have an rms variation of (0.11, 0.09) mas and
a mean offset (0.06, 0.22) mas from the astrometric position obtained in paper I.
That mean offset is within the error bars of the original measurement.
These measures for the secondary calibrator provide an estimate of the
astrometric accuracy of the experiment.  

We summarize positions and flux densities for GJ 896A in Table~\ref{tab:data} as a 
function of epoch.  We also include measurements from Paper I.  GJ 896A is strongly
variable:  flux densities vary from 0.140 to 9.512 mJy during the course of the
experiment.  The images of GJ 896A are predominantly point-like, as found in Paper I.
The smaller errors on the flux density and the positions are reflective
of the increased sensitivity and $(u,v)$ coverage of RIPL over the precursor 
experiment due to the doubling of the observing time, doubling of the bit rate, and
addition of the GBT to the array.  The statistical errors are significantly less
than the systematic errors associated with the astrometry.  For the analysis
discussed below, we add in quadrature an error of 0.2 mas to each coordinate as
an estimate of the systematic error.

\section{Astrometric Fitting}

In addition to the eight epochs from RIPL, we obtained three epochs in the RIPL precursor
survey over three days in March 2006.  We consider the data from all 11 epochs in 
our astrometric fitting here.  The data span 4.9 years.

GJ 896A was not originally part of the Hipparcos catalog but it is included in
the reanalysis of the Hipparcos data \citep{2007A&A...474..653V}.
We summarize the known astrometric parameters in
Table~\ref{tab:astro}.  GJ 896A is also known to be in
a double system with GJ 896B; parameters are determined from $\sim 40$ years of optical observations
\cite[Table~\ref{tab:binary};][]{1984AJ.....89.1063H,2001AJ....122.3466M}.  
Since, the baseline for this orbit determination 
is $\sim 0.1$ times the estimated period, the orbital parameters are poorly determined.
This is reflected in the binary orbit grade of 5 (``indeterminate'') on a scale of 1 to 5.

\begin{figure}
\includegraphics[width=1.0\textwidth]{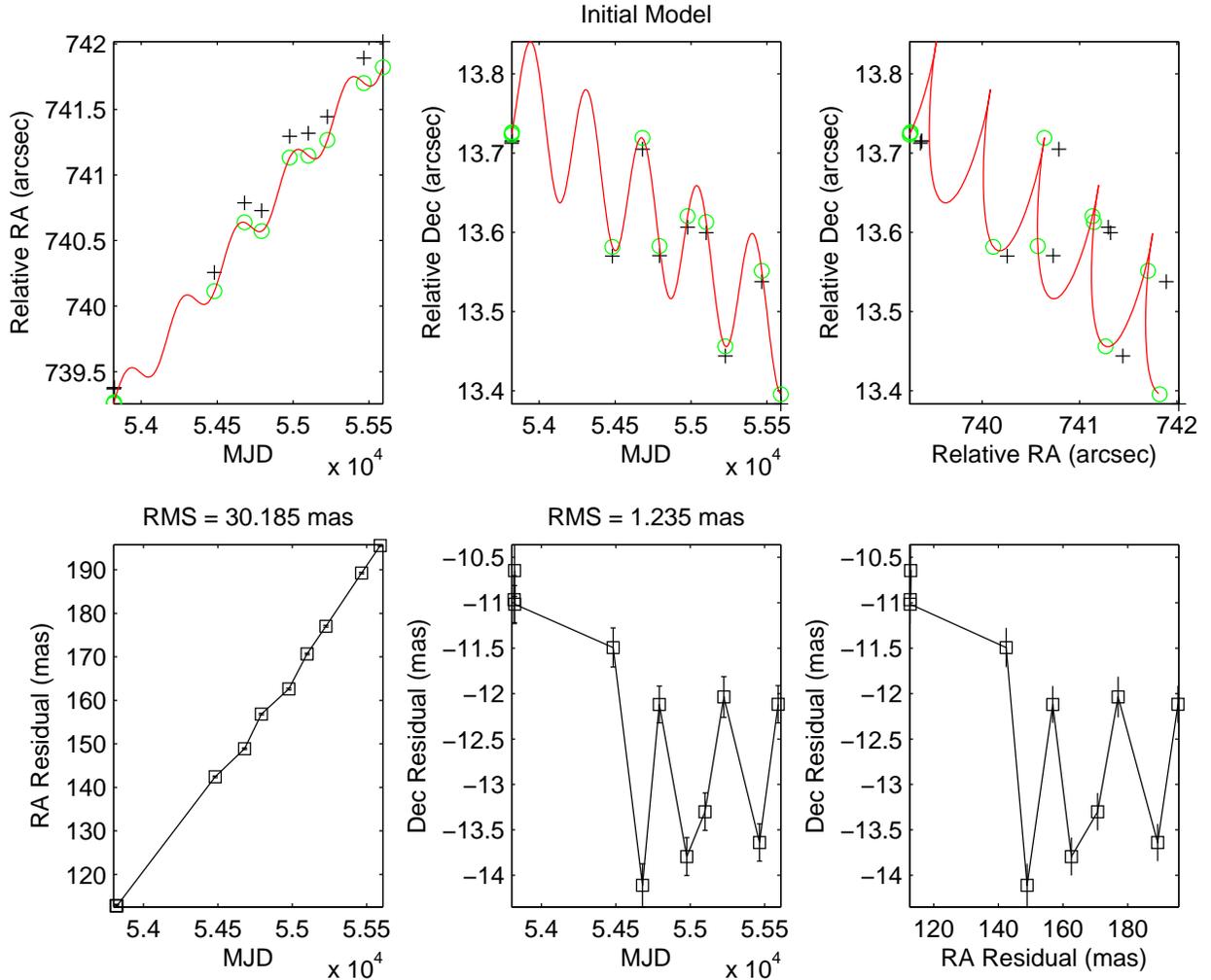}
\caption{Initial model and astrometric data for GJ 896A.  Top row shows 
astrometric data (black crosses) with the initial (optically-determined) astrometric 
model (green circles at the epoch of observation and red lines for the continuous 
calculation).  Positions in the top row are given in arcsec relative to a fiducial
position of 23$^h$31$^m$, 19$^\circ$56'.
Bottom row shows residuals after subtraction of the model.
The symbols in the top row are much larger than the errors; in the bottom
row, errors include statistical and systematic contributions.
\label{fig:initial}  
}
\end{figure}

\begin{figure}
\includegraphics[width=1.0\textwidth]{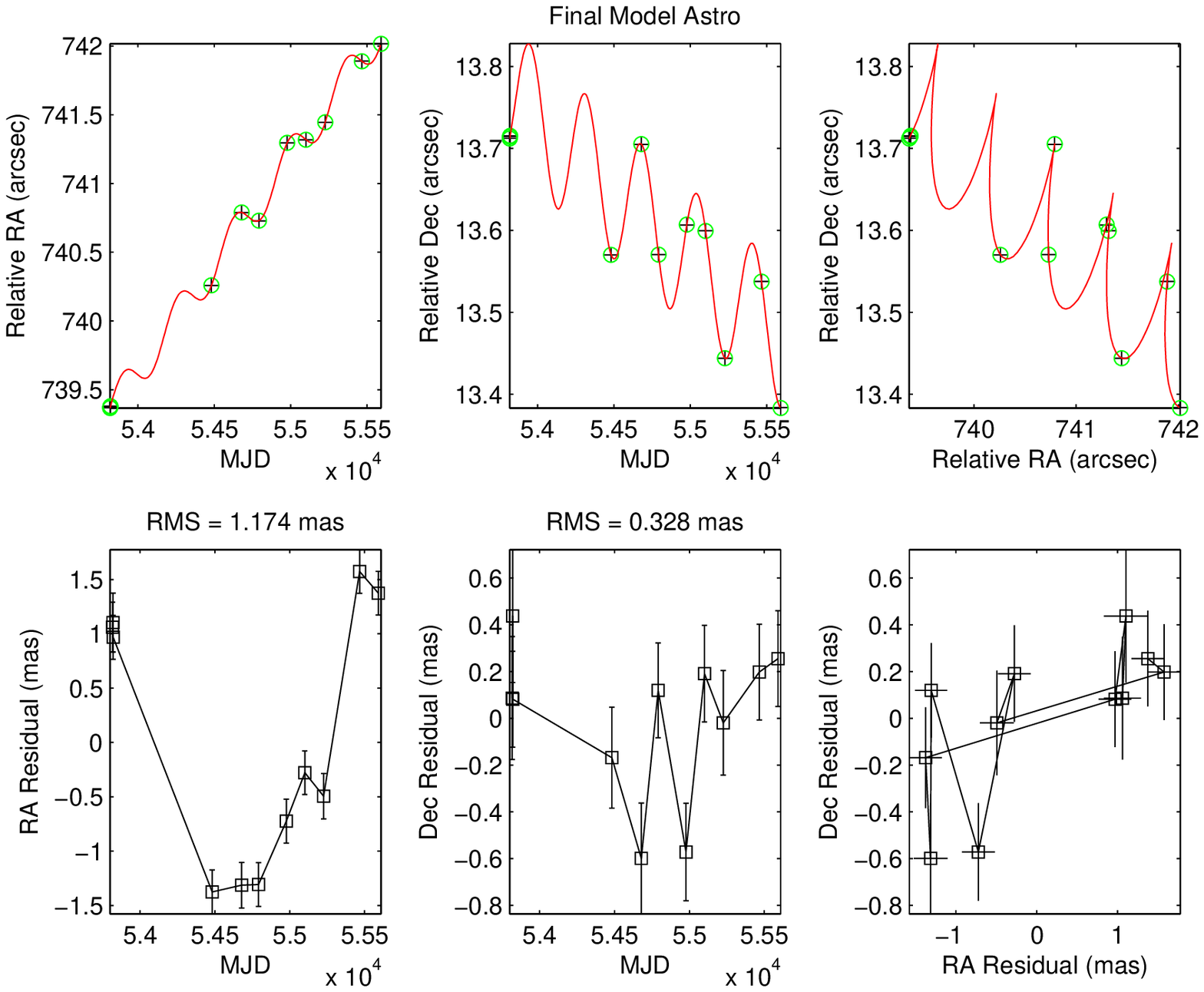}
\caption{Fitted astrometric model and astrometric data for GJ 896A.  
Symbols are the same as in Figure~\ref{fig:initial}.
\label{fig:astro}
}
\end{figure}

\begin{figure}
\includegraphics[width=1.0\textwidth]{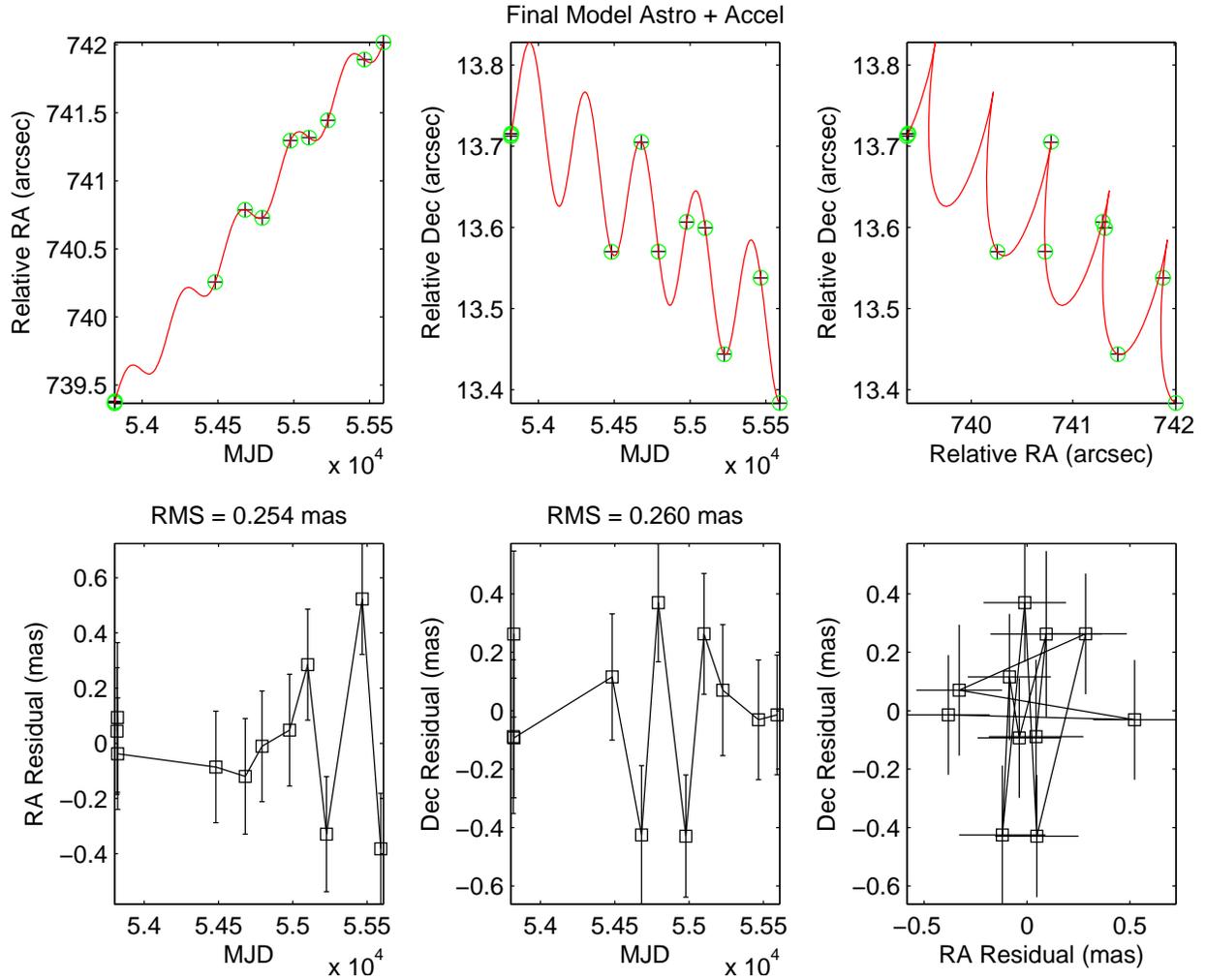}
\caption{Fitted astrometric plus acceleration model and astrometric data for GJ 896A.  
Symbols are the same as in Figure~\ref{fig:initial}.
\label{fig:accel}
}
\end{figure}

In Table~\ref{tab:astro}, we show 
the nominal astrometric data available from optical sources 
as well as best-fit solutions based on RIPL data.  We performed two kinds of fits:
one with just the astrometric parameters, and the other with astrometric parameters
and an acceleration term.  
Figures~\ref{fig:initial},~\ref{fig:astro}, and~\ref{fig:accel} show the astrometric
data, the models, and the residuals for the cases of the initial model, the astrometric
model, and the astrometric plus acceleration model, respectively.  The astrometric
plus acceleration model that we fit to our data $(\alpha(t_i),\delta(t_i))$ at epoch $t_i$ is
\begin{eqnarray}
\alpha(t_i) &=& \alpha_0 + \mu_\alpha  (t_i - t_0) + \pi_\alpha (\alpha, \delta, t_i) + a_\alpha  (t_i - t_0^\alpha)^2, \\
\delta(t_i) &=& \delta_0 + \mu_\delta  (t_i - t_0) + \pi_\delta (\alpha, \delta, t_i) + a_\delta  (t_i - t_0^\delta)^2, 
\end{eqnarray}
where $(\alpha_0,\delta_0)$ is the equinox J2000 position at time $t_0$, the epoch 2000.0, 
$(\mu_\alpha, \mu_\delta)$ is the proper motion, $(\pi_\alpha,\pi_\delta)$ are the projections of the parallax
$\pi$ onto the two coordinates,
($a_\alpha$, $a_\delta$) is the local acceleration, and $t_0^\alpha$ and $t_0^\delta$ are zero-points for the
acceleration in each coordinate.  
The initial model clearly shows a substantial error in $\mu_\alpha$ and a $\sim 1$ mas 
error in $\pi$.  Fitting 
the astrometric parameters ($\alpha_0$, $\delta_0$, $\mu_\alpha$, $\mu_\delta$, $\pi$)
substantially improves the quality of the fit but leaves a parabolic residual in
$\alpha$.  An acceleration term reduces the rms in $\alpha$ from $1.2$ to 0.26 mas.
The residual in declination is improved from 0.33 to 0.26 mas through the addition
of an acceleration term.  The astrometric fit alone with 22 measurements
and 5 parameters producing 17 degrees of freedom results in $\chi^2=338$.
The acceleration term significantly reduces $\chi^2$ to 29.7 
with 9 parameters producing 13 degrees of freedom.  The reduced $\chi^2_\nu=2.3$
in this final fit.

\begin{figure}
\includegraphics[width=1.0\textwidth]{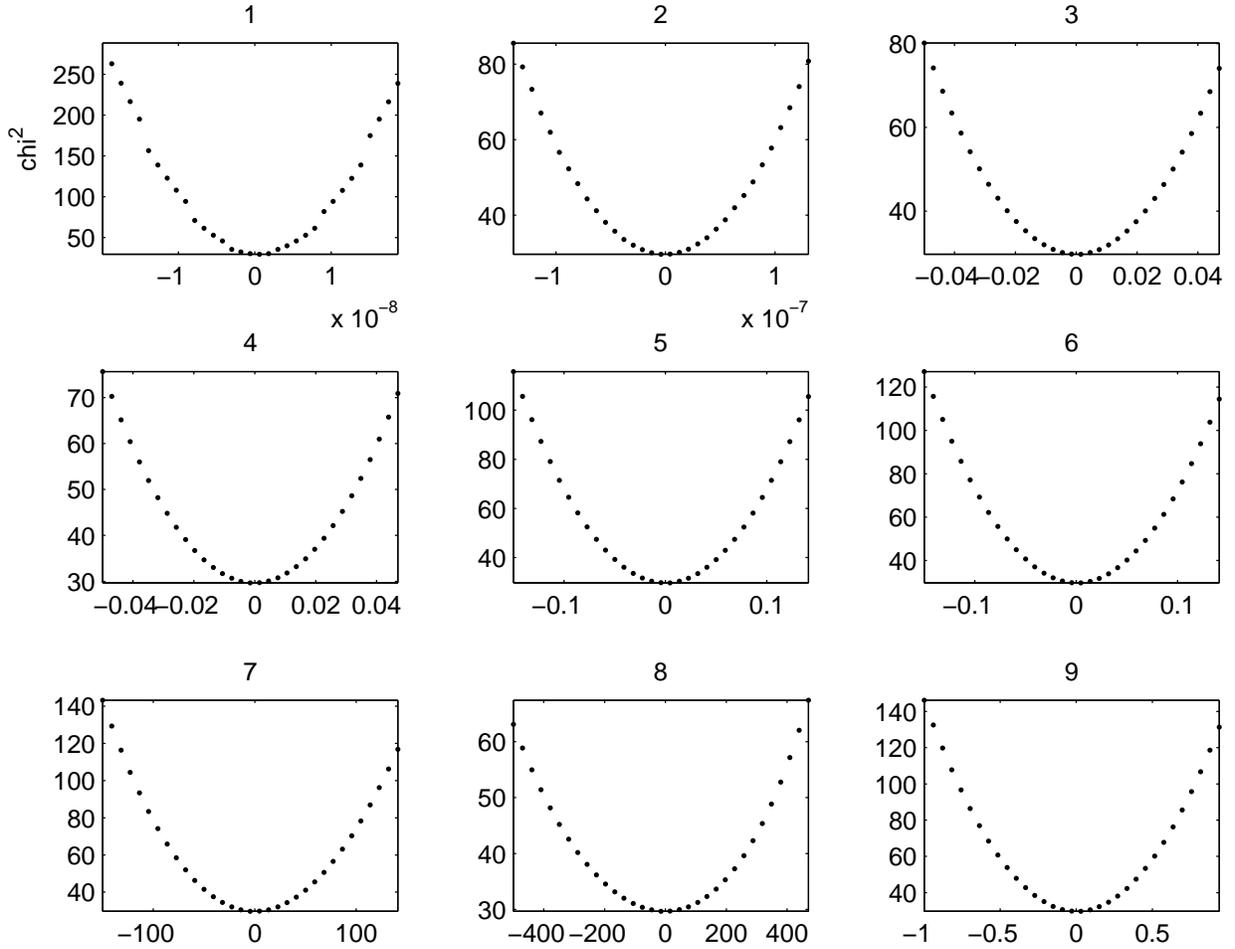}
\caption{$\chi^2$ as a function of each parameter.  The x-axis values are offsets
from the best-fit value.  The parameters explored are (1) right ascension (hours),
(2) declination (degrees), (3) proper motion in right ascension (mas/y), (4) proper motion
in declination (mas/y), (5) acceleration in right ascension (mas/y$^2$), (6) acceleration in
declination (mas/y$^2$), (7) minimum time for the right ascension acceleration (days), 
(8) maximum time for the declination acceleration (days), and (9) parallax (mas).
\label{fig:chi2}
}
\end{figure}

We estimate errors in individual parameters by mapping the $\chi^2$ surface as a function 
of each parameter.  For the purely astrometric case, the $\chi^2$ curves are parabolic.
We plot the $\chi^2$ curves for the astrometric and acceleration terms in Figure~\ref{fig:chi2}.
These curves are also parabolic and symmetric about their best-fit values.  
We estimate the $1\sigma$ errors from these curves.
The astrometric parameter estimates are significantly more accurate when the acceleration
terms are included.  We do not estimate the full covariance matrix of the parameter estimates due
to the computational cost of exploring a 9-dimensional space.  We did  explore
the covariance between the parallax and the right ascension acceleration term, two parameters
that one might expect to be strongly linked.  We find that the $\chi^2$ surface in this
two-dimensional domain is well-characterized by a paraboloidal surface with axes that are misaligned
from the principle axes by less than 30 degrees; that is, the parameters are essentially independent.

The estimated values for the astrometric parameters are determined one to two orders of
magnitude more accurately than the Hipparcos values.  The values for 
$\alpha_0$, $\delta_0$, $\pi$, and $\mu_\delta$ are in agreement with the Hipparcos
values.  The proper motion in right ascension, $\mu_\alpha$, however, differs 
by $16.5 \pm 1.4$ mas/y.  Over the time baseline between the Hipparcos epoch (1991.25)
and the mean RIPL epoch (2008.75), this implies an acceleration of $0.94 \pm 0.08$ mas/y$^2$.
This has the same sign and a comparable magnitude to the acceleration term measured
from the RIPL data alone.  The agreement in $\mu_\delta$ implies an upper limit of 0.06
mas/y$^2$ over this interval.

Motion of the star during the observing epoch can be an important effect, especially if
the flux density is variable on short timescales.  For GJ 896A, the median absolute value of motions from astrometric
parameters are (0.10, 0.048) mas/hour in the two coordinates, respectively,
and are always lower than (0.14, 0.065) mas/hour.  
This can lead to smearing out of the source 
and reduction of the peak flux density.  If the source flux density is strongly variable during the observation,
then the mean time of the epoch may not correspond to the time of the astrometric detection, leading  
to errors be as large as (0.6,0.3) mas.  The small residuals we obtain from fitting indicate that 
this effect is not likely to be important in most epochs.  Future reductions of this data will attempt
to determine any light curve variability and its effect on the astrometry and/or include
corrections to the visibilities for the assumed astrometric motion of the star.  In high signal to noise ratio
cases, we may be able to simultaneously fit a position and velocity within a given epoch.

\section{Constraints on Companions to GJ 896A}

\begin{figure}
\includegraphics[width=1.0\textwidth]{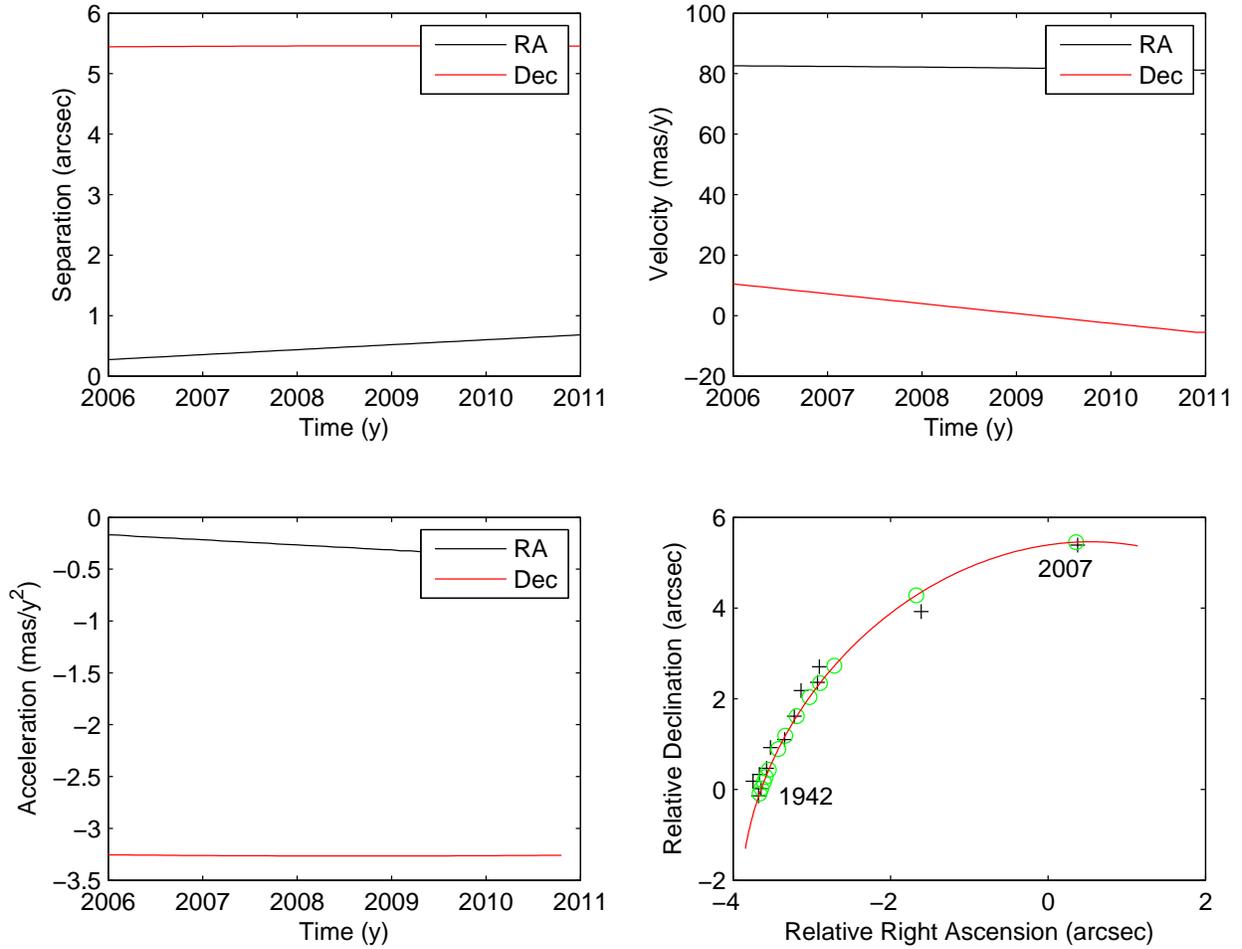}
\caption{Orbital motion for GJ 896B relative to GJ 896A based on optical
astrometric data.  We plot the relative separation, velocity, and acceleration
in the upper left, upper right, and lower left, respectively.  We plot
in the lower right the historical data (black crosses) from which this orbit derived 
\citep{1984AJ.....89.1063H} and the
predictions of the orbital model (green circles at the time of measurement, red line for 
continuous time).  We include an additional datum
from 2007 that was not part of the fit
\citep{2001AJ....122.3466M}.  The plotted model curve extends from 1932 to 2017.
\label{fig:binary}
}
\end{figure}

The best fit accelerations for GJ 896A are (0.458, 0.087) mas/y$^2$.  We compare
this to the known parameters of the binary system.  We plot relative
separation, velocity, and acceleration determined from the binary parameters
(Figure~\ref{fig:binary}).  Note that the definition of the orbit gives
the position of B relative to A.  Thus, the sign of velocity and
acceleration terms is inverted relative to our measurement.  If the stars
have equal mass, then the motion relative to the center of mass will be half
of the value plotted.
The mean calculated accelerations for A relative to B over our observing epochs is 
($0.3 \pm 0.1$, $3.1 \pm 0.6$) mas/y$^2$.
The calculated acceleration term in $\alpha$ agrees reasonably
well with the measured value and is comparable to 
the mean acceleration over the entire orbit $\approx 2\pi r / P^2 =0.3 $ mas/y$^2$.
The calculated acceleration term $a_\delta$ does not agree at all with the measured
value.  The RIPL measurements occur near the predicted periastron, and, therefore,
a maximum in the value of $a_\delta$.  Our results clearly indicate a significant
error in the model for this binary system and appear to indicate that the period
may be longer than 359 years, or that the eccentricity or time of periastron have not been 
properly estimated.

The acceleration $a$ allows us to a place a limit on the stellar companion mass, $M_B$,
using
\begin{equation}
\left( M_B \over M_J \right) = \left( a \over 0.0381 {\rm\ AU/y^2} \right) \left( r \over 1 {\rm AU} \right)^2,
\end{equation}
where $r$ is the current separation projected onto the sky and $M_J$ is the mass of Jupiter.
If we translate the measured acceleration to physical units, we find $a_\alpha = 2.8\times 10^{-3}$
AU/y$^2$ and $a_\delta=0.5\times 10^{-3}$ AU/y$^2$, or the total acceleration $a =2.9\times 10^{-3}$ 
AU/y$^2$.  Applied to the stellar companion at a projected distance of 5.5\arcsec 
(the optically measured separation in 2007), we estimate a
mass for GJ 896B of $M_B \approx 0.4 M_\odot$.  This mass estimate is in rough agreement
with expectations for a star of spectral type M4.5 \citep{2011ApJ...728...48K}.

\begin{figure}
\includegraphics[width=1.0\textwidth]{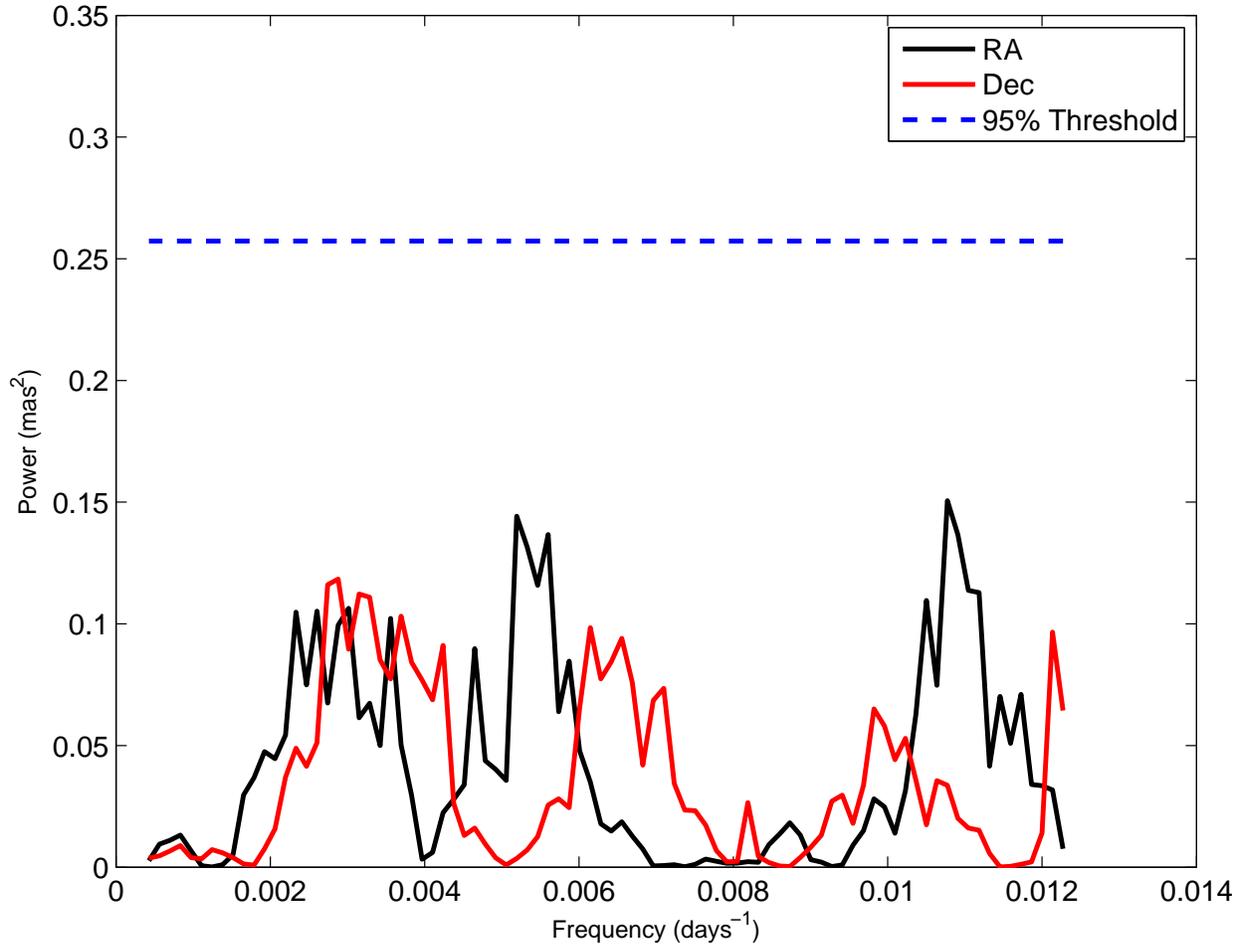}
\caption{Periodogram power as a function of frequency for residuals after
removal of astrometric and acceleration terms for the two coordinates.
We also plot the 95\% confidence threshold as the blue dashed line.
The absence of a significant peak in the periodogram places limits 
on short-period companions.
\label{fig:lomb}
}
\end{figure}

\begin{figure}
\includegraphics[width=1.0\textwidth]{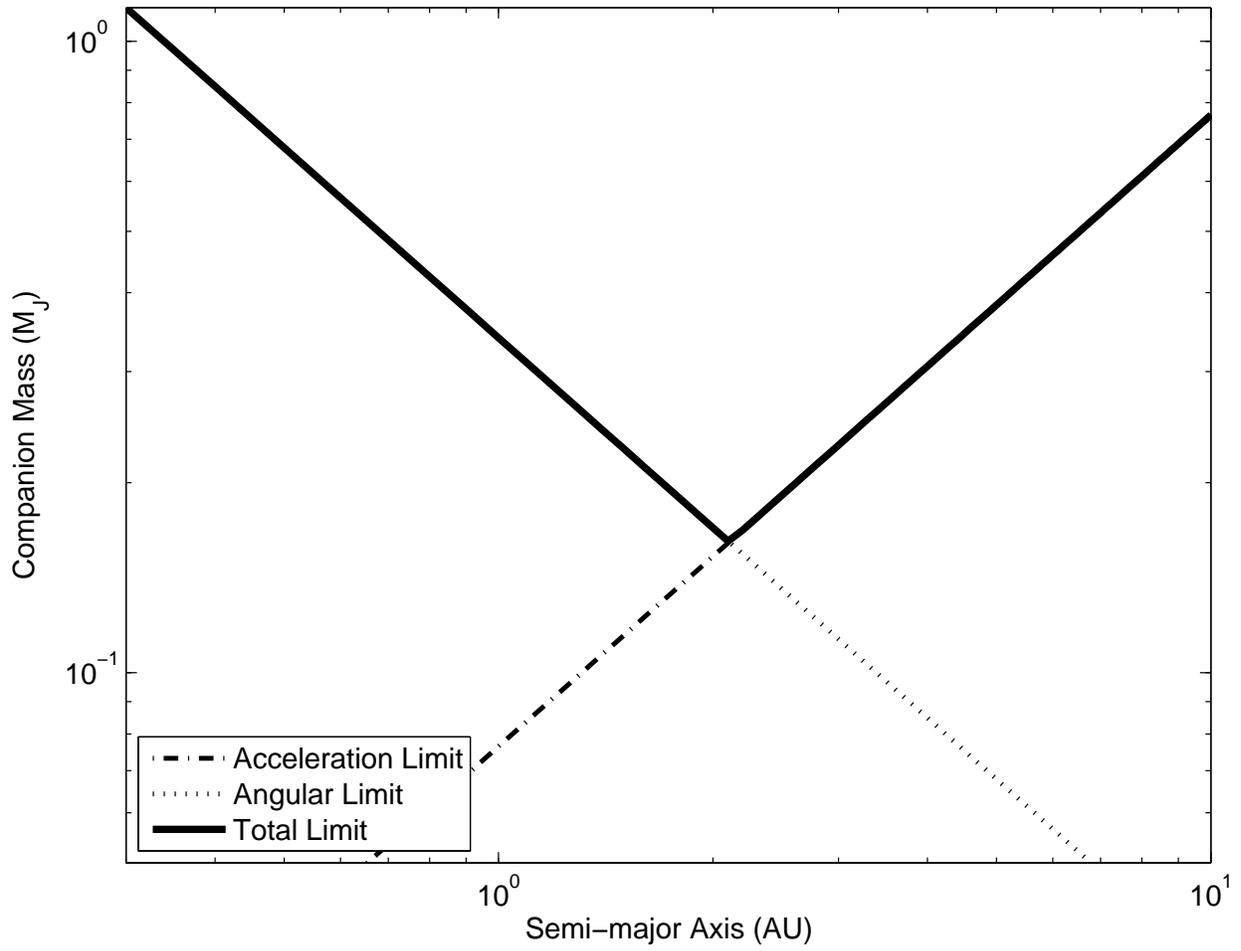}
\caption{Limits on companion mass as a function of semi-major axis for GJ 896A.
Limits are derived from acceleration and angular separation terms.
The space below the solid curve is allowed.
\label{fig:limits}
}
\end{figure}

We can also apply the acceleration as a limit to estimate planetary mass companions.
If we take the acceleration $a$ as an upper limit then we can place a limit on a planetary mass,  
$M_p \lsim 0.08 M_J \left(r /1AU \right)^2$.  This acceleration limit applies to companions 
with a period $\gsim 4.9$ y, or at $r \gsim 1.7$ AU for a stellar mass of 0.2 $M_\odot$.
For closer orbits, we can set a constraint based on the maximal amplitude of angular 
displacement.  We set this limit by computing the periodogram for the residuals in
each coordinate after subtraction of the astrometric plus acceleration model (Figure~\ref{fig:lomb}).
The amplitude of the periodogram falls well below the 95\%-confidence threshold; we find
no evidence for a short-period companion.  The maximal power in the periodogram corresponds to 
an upper limit of 0.4 mas in angular displacement.  We are sensitive to orbital periods as
short as the characteristic separation between epochs $\sim 140$ days; this limit translates
to an inner radius cutoff of $\sim 0.3$ AU.  In Figure~\ref{fig:limits}, we combine
the acceleration limits and the angular displacement limits.  The maximal sensitivity
is at  2 AU with an upper limit of 0.15 $M_J$.  Thus, we are sensitive to a planet with 
approximately 3 Neptune masses at large radii from GJ 896A.  Note that this limit 
on the mass applies strictly to face-on systems; orbits at all inclinations will have a
maximum acceleration equal to the face-on case but will have smaller instantaneous values 
depending on the phase of the orbit.

A caveat to this calculation is that the acceleration we measure may not be a upper limit
to the effect of a planetary companion if the companion and GJ 896B produce accelerations
that almost cancel each other.  The agreement in $a_\alpha$ between measured and
calculated values suggests that this is not likely; the disagreement in $a_\delta$ 
could be due to a planetary companion acceleration that wholly or partially
cancels the binary acceleration.  A Jupiter mass companion at 2 AU would produce 
the magnitude of acceleration necessary for the cancellation.
This cancellation, however, requires a coincidence that is unlikely to persist over a period of 
observation of this duration.  The agreement in $\mu_\delta$ between the 
Hipparcos and RIPL measurements indicates that such a cancellation must persist over a period
of $\gsim 16$ y, which is improbable.  Improved optical astrometry of the binary system can determine
the contribution of GJ 896B to the motion of GJ 896A.

The rms residuals indicate that the centroid of the stellar emission is confined to a physical
region $\lsim 0.3 R_\odot$.  This is comparable to or less than the photospheric radius of a late-type star
\citep[e.g.,][]{2011ApJ...728...48K}.  These residuals occur over a range of source flux densities that vary by a factor of
$\sim 60$.  We conclude that the magnetic activity associated with GJ 896A is symmetric about
the stellar surface and/or confined primarily to the disk of the star.

\section{Conclusions}

We have presented here an analysis of RIPL observations of the star GJ 896A.  We measure the 
parallax to an accuracy of 0.2 mas and the proper motion to an accuracy of 0.01 mas/y.  We
measure an acceleration term of $0.466 \pm 0.045$ mas/y$^2$.  This term is significantly less
than expected based on the model for the orbit of the GJ 896AB system, suggesting an error in the 
optically-determined orbital parameters.  Attributing this acceleration to a planetary
mass companion, we demonstrate sensitivity to a planetary companion at 2 AU of mass 0.15 $M_J$.

The residuals that we obtain from our fitting of 0.25 mas in each coordinate over 4.9 y indicate
the feasibility of planet searching and characterization with radio astrometry of nearby low mass
stars.  These residuals can potentially be improved for this star and others through 
technical improvements to the data analysis including accounting for the motion of the star during
the observing epoch and more exact removal of tropospheric fluctuations.  The residuals 
indicate that astrometric jitter from stellar activity on scales of a stellar radius or larger
is not a limiting factor in planet astrometry.  Improvements in our analysis of systematic
errors from stellar motion and tropospheric phase errors will permit us to explore
sub-stellar-radius astrometric accuracy.

Future RIPL papers will explore analysis of additional data from this star,
improved analysis techniques, and analysis of other stars in the sample.  Upgrades to the 
recording
bandwidth of the VLBA (and GBT) will produce significant improvements in the capability to carry out
observing programs of this kind.  The improved sensitivity will allow more rapid detection of
bright stars such as GJ 896A, thereby reducing any smearing effect from stellar motion,
allow the use of closer calibrators for improved tropospheric corrections and reduced
time on calibrator, provide higher signal to noise ratio detections of fainter stars, and 
enable us to detect a larger sample of stars.

\acknowledgements
The National Radio Astronomy Observatory is a facility of the National Science Foundation operated under cooperative agreement by Associated Universities, Inc.
This research has made use of the NASA/IPAC
Extragalactic Database (NED) which is operated by the Jet Propulsion
Laboratory, California Institute of Technology, under contract with
the National Aeronautics and Space Administration. This research has
made use of the SIMBAD database, operated at CDS, Strasbourg, France.
This research was supported by NASA Origins grant NNX07AP19G.

%\bibliographystyle{apj-hacked}
%\bibliography{myrefs}

\begin{thebibliography}{36}
\expandafter\ifx\csname natexlab\endcsname\relax\def\natexlab#1{#1}\fi

\bibitem[{{Beaulieu} {et~al.}(2006){Beaulieu}, {Bennett}, {Fouqu{\'e}},
  {Williams}, {Dominik}, {J{\o}rgensen}, {Kubas}, {Cassan}, {Coutures},
  {Greenhill}, {Hill}, {Menzies}, {Sackett}, {Albrow}, {Brillant}, {Caldwell},
  {Calitz}, {Cook}, {Corrales}, {Desort}, {Dieters}, {Dominis}, {Donatowicz},
  {Hoffman}, {Kane}, {Marquette}, {Martin}, {Meintjes}, {Pollard}, {Sahu},
  {Vinter}, {Wambsganss}, {Woller}, {Horne}, {Steele}, {Bramich}, {Burgdorf},
  {Snodgrass}, {Bode}, {Udalski}, {Szyma{\'n}ski}, {Kubiak}, {Wi{\c e}ckowski},
  {Pietrzy{\'n}ski}, {Soszy{\'n}ski}, {Szewczyk}, {Wyrzykowski},
  {Paczy{\'n}ski}, {Abe}, {Bond}, {Britton}, {Gilmore}, {Hearnshaw}, {Itow},
  {Kamiya}, {Kilmartin}, {Korpela}, {Masuda}, {Matsubara}, {Motomura},
  {Muraki}, {Nakamura}, {Okada}, {Ohnishi}, {Rattenbury}, {Sako}, {Sato},
  {Sasaki}, {Sekiguchi}, {Sullivan}, {Tristram}, {Yock}, \&
  {Yoshioka}}]{2006Natur.439..437B}
{Beaulieu}, J.-P., {Bennett}, D.~P., {Fouqu{\'e}}, P., {et~al.} 2006, \nat,
  439, 437

\bibitem[{{Bennett} {et~al.}(2008){Bennett}, {Bond}, {Udalski}, {Sumi}, {Abe},
  {Fukui}, {Furusawa}, {Hearnshaw}, {Holderness}, {Itow}, {Kamiya}, {Korpela},
  {Kilmartin}, {Lin}, {Ling}, {Masuda}, {Matsubara}, {Miyake}, {Muraki},
  {Nagaya}, {Okumura}, {Ohnishi}, {Perrott}, {Rattenbury}, {Sako}, {Saito},
  {Sato}, {Skuljan}, {Sullivan}, {Sweatman}, {Tristram}, {Yock}, {Kubiak},
  {Szyma{\'n}ski}, {Pietrzy{\'n}ski}, {Soszy{\'n}ski}, {Szewczyk},
  {Wyrzykowski}, {Ulaczyk}, {Batista}, {Beaulieu}, {Brillant}, {Cassan},
  {Fouqu{\'e}}, {Kervella}, {Kubas}, \& {Marquette}}]{2008ApJ...684..663B}
{Bennett}, D.~P., {Bond}, I.~A., {Udalski}, A., {et~al.} 2008, \apj, 684, 663

\bibitem[{{Benz} {et~al.}(1995){Benz}, {Alef}, \&
  {Guedel}}]{1995A&A...298..187B}
{Benz}, A.~O., {Alef}, W., \& {Guedel}, M. 1995, \aap, 298, 187

\bibitem[{{Borucki} {et~al.}(2011){Borucki}, {Koch}, {Basri}, {Batalha},
  {Boss}, {Brown}, {Caldwell}, {Christensen-Dalsgaard}, {Cochran}, {DeVore},
  {Dunham}, {Dupree}, {Gautier}, {Geary}, {Gilliland}, {Gould}, {Howell},
  {Jenkins}, {Kjeldsen}, {Latham}, {Lissauer}, {Marcy}, {Monet}, {Sasselov},
  {Tarter}, {Charbonneau}, {Doyle}, {Ford}, {Fortney}, {Holman}, {Seager},
  {Steffen}, {Welsh}, {Allen}, {Bryson}, {Buchhave}, {Chandrasekaran},
  {Christiansen}, {Ciardi}, {Clarke}, {Dotson}, {Endl}, {Fischer}, {Fressin},
  {Haas}, {Horch}, {Howard}, {Isaacson}, {Kolodziejczak}, {Li}, {MacQueen},
  {Meibom}, {Prsa}, {Quintana}, {Rowe}, {Sherry}, {Tenenbaum}, {Torres},
  {Twicken}, {Van Cleve}, {Walkowicz}, \& {Wu}}]{2011ApJ...728..117B}
{Borucki}, W.~J., {Koch}, D.~G., {Basri}, G., {et~al.} 2011, \apj, 728, 117

\bibitem[{{Boss}(2006)}]{2006ApJ...643..501B}
{Boss}, A.~P. 2006, \apj, 643, 501

\bibitem[{{Boss} {et~al.}(2009){Boss}, {Weinberger}, {Anglada-Escud{\'e}},
  {Thompson}, {Burley}, {Birk}, {Pravdo}, {Shaklan}, {Gatewood}, {Majewski}, \&
  {Patterson}}]{2009PASP..121.1218B}
{Boss}, A.~P., {Weinberger}, A.~J., {Anglada-Escud{\'e}}, G., {et~al.} 2009,
  \pasp, 121, 1218

\bibitem[{{Bower} {et~al.}(2007){Bower}, {Bolatto}, {Ford}, {Kalas}, \&
  {Ulvestad}}]{2007arXiv0704.0238B}
{Bower}, G.~C., {Bolatto}, A., {Ford}, E., {Kalas}, P., \& {Ulvestad}, J. 2007,
  ArXiv e-prints

\bibitem[{{Bower} {et~al.}(2009){Bower}, {Bolatto}, {Ford}, \&
  {Kalas}}]{2009ApJ...701.1922B}
{Bower}, G.~C., {Bolatto}, A., {Ford}, E.~B., \& {Kalas}, P. 2009, \apj, 701,
  1922

\bibitem[{{Brunthaler} {et~al.}(2011){Brunthaler}, {Reid}, {Menten}, {Zheng},
  {Bartkiewicz}, {Choi}, {Dame}, {Hachisuka}, {Immer}, {Moellenbrock},
  {Moscadelli}, {Rygl}, {Sanna}, {Sato}, {Wu}, {Xu}, \&
  {Zhang}}]{2011AN....332..461B}
{Brunthaler}, A., {Reid}, M.~J., {Menten}, K.~M., {et~al.} 2011, Astronomische
  Nachrichten, 332, 461

\bibitem[{{Butler} {et~al.}(2006){Butler}, {Wright}, {Marcy}, {Fischer},
  {Vogt}, {Tinney}, {Jones}, {Carter}, {Johnson}, {McCarthy}, \&
  {Penny}}]{2006ApJ...646..505B}
{Butler}, R.~P., {Wright}, J.~T., {Marcy}, G.~W., {et~al.} 2006, \apj, 646, 505

\bibitem[{{Endl} {et~al.}(2006){Endl}, {Cochran}, {K{\"u}rster}, {Paulson},
  {Wittenmyer}, {MacQueen}, \& {Tull}}]{2006ApJ...649..436E}
{Endl}, M., {Cochran}, W.~D., {K{\"u}rster}, M., {et~al.} 2006, \apj, 649, 436

\bibitem[{{Fomalont} \& {Kopeikin}(2003)}]{2003ApJ...598..704F}
{Fomalont}, E.~B., \& {Kopeikin}, S.~M. 2003, \apj, 598, 704

\bibitem[{{Forbrich} \& {Berger}(2009)}]{2009ApJ...706L.205F}
{Forbrich}, J., \& {Berger}, E. 2009, \apjl, 706, L205

\bibitem[{{Gagne} {et~al.}(1998){Gagne}, {Valenti}, {Johns-Krull}, {Linsky},
  {Brown}, \& {Gudel}}]{1998ASPC..154.1484G}
{Gagne}, M., {Valenti}, J., {Johns-Krull}, C., {et~al.} 1998, in Astronomical
  Society of the Pacific Conference Series, Vol. 154, Cool Stars, Stellar
  Systems, and the Sun, ed. {R.~A.~Donahue \& J.~A.~Bookbinder}, 1484--+

\bibitem[{{Gaudi} {et~al.}(2008){Gaudi}, {Bennett}, {Udalski}, {Gould},
  {Christie}, {Maoz}, {Dong}, {McCormick}, {Szyma{\'n}ski}, {Tristram},
  {Nikolaev}, {Paczy{\'n}ski}, {Kubiak}, {Pietrzy{\'n}ski}, {Soszy{\'n}ski},
  {Szewczyk}, {Ulaczyk}, {Wyrzykowski}, {DePoy}, {Han}, {Kaspi}, {Lee},
  {Mallia}, {Natusch}, {Pogge}, {Park}, {Abe}, {Bond}, {Botzler}, {Fukui},
  {Hearnshaw}, {Itow}, {Kamiya}, {Korpela}, {Kilmartin}, {Lin}, {Masuda},
  {Matsubara}, {Motomura}, {Muraki}, {Nakamura}, {Okumura}, {Ohnishi},
  {Rattenbury}, {Sako}, {Saito}, {Sato}, {Skuljan}, {Sullivan}, {Sumi},
  {Sweatman}, {Yock}, {Albrow}, {Allan}, {Beaulieu}, {Burgdorf}, {Cook},
  {Coutures}, {Dominik}, {Dieters}, {Fouqu{\'e}}, {Greenhill}, {Horne},
  {Steele}, {Tsapras}, {Chaboyer}, {Crocker}, {Frank}, \&
  {Macintosh}}]{2008Sci...319..927G}
{Gaudi}, B.~S., {Bennett}, D.~P., {Udalski}, A., {et~al.} 2008, Science, 319,
  927

\bibitem[{{Gould} {et~al.}(2006){Gould}, {Udalski}, {An}, {Bennett}, {Zhou},
  {Dong}, {Rattenbury}, {Gaudi}, {Yock}, {Bond}, {Christie}, {Horne},
  {Anderson}, {Stanek}, {DePoy}, {Han}, {McCormick}, {Park}, {Pogge},
  {Poindexter}, {Soszy{\'n}ski}, {Szyma{\'n}ski}, {Kubiak}, {Pietrzy{\'n}ski},
  {Szewczyk}, {Wyrzykowski}, {Ulaczyk}, {Paczy{\'n}ski}, {Bramich},
  {Snodgrass}, {Steele}, {Burgdorf}, {Bode}, {Botzler}, {Mao}, \&
  {Swaving}}]{2006ApJ...644L..37G}
{Gould}, A., {Udalski}, A., {An}, D., {et~al.} 2006, \apjl, 644, L37

\bibitem[{{Greisen}(2003)}]{2003ASSL..285..109G}
{Greisen}, E.~W. 2003, in Astrophysics and Space Science Library, Vol. 285,
  Astrophysics and Space Science Library, ed. A.~{Heck}, 109

\bibitem[{{Guirado} {et~al.}(2006){Guirado}, {Mart{\'{\i}}-Vidal}, {Marcaide},
  {Close}, {Algaba}, {Brandner}, {Lestrade}, {Jauncey}, {Jones}, {Preston}, \&
  {Reynolds}}]{2006A&A...446..733G}
{Guirado}, J.~C., {Mart{\'{\i}}-Vidal}, I., {Marcaide}, J.~M., {et~al.} 2006,
  \aap, 446, 733

\bibitem[{{Heintz}(1984)}]{1984AJ.....89.1063H}
{Heintz}, W.~D. 1984, \aj, 89, 1063

\bibitem[{{Ida} \& {Lin}(2005)}]{2005ApJ...626.1045I}
{Ida}, S., \& {Lin}, D.~N.~C. 2005, \apj, 626, 1045

\bibitem[{{Jackson} {et~al.}(1989){Jackson}, {Kundu}, \&
  {White}}]{1989A&A...210..284J}
{Jackson}, P.~D., {Kundu}, M.~R., \& {White}, S.~M. 1989, \aap, 210, 284

\bibitem[{{Johnson} {et~al.}(2007){Johnson}, {Butler}, {Marcy}, {Fischer},
  {Vogt}, {Wright}, \& {Peek}}]{2007ApJ...670..833J}
{Johnson}, J.~A., {Butler}, R.~P., {Marcy}, G.~W., {et~al.} 2007, \apj, 670,
  833

\bibitem[{{Kennedy} {et~al.}(2007){Kennedy}, {Kenyon}, \&
  {Bromley}}]{2007Ap&SS.311....9K}
{Kennedy}, G.~M., {Kenyon}, S.~J., \& {Bromley}, B.~C. 2007, \apss, 311, 9

\bibitem[{{Kraus} {et~al.}(2011){Kraus}, {Tucker}, {Thompson}, {Craine}, \&
  {Hillenbrand}}]{2011ApJ...728...48K}
{Kraus}, A.~L., {Tucker}, R.~A., {Thompson}, M.~I., {Craine}, E.~R., \&
  {Hillenbrand}, L.~A. 2011, \apj, 728, 48

\bibitem[{{Laughlin} {et~al.}(2004){Laughlin}, {Bodenheimer}, \&
  {Adams}}]{2004ApJ...612L..73L}
{Laughlin}, G., {Bodenheimer}, P., \& {Adams}, F.~C. 2004, \apjl, 612, L73

\bibitem[{{Lestrade} {et~al.}(1999){Lestrade}, {Preston}, {Jones}, {Phillips},
  {Rogers}, {Titus}, {Rioja}, \& {Gabuzda}}]{1999A&A...344.1014L}
{Lestrade}, J.-F., {Preston}, R.~A., {Jones}, D.~L., {et~al.} 1999, \aap, 344,
  1014

\bibitem[{{Loinard} {et~al.}(2007){Loinard}, {Torres}, {Mioduszewski},
  {Rodr{\'{\i}}guez}, {Gonz{\'a}lez-L{\'o}pezlira}, {Lachaume}, {V{\'a}zquez},
  \& {Gonz{\'a}lez}}]{2007ApJ...671..546L}
{Loinard}, L., {Torres}, R.~M., {Mioduszewski}, A.~J., {et~al.} 2007, \apj,
  671, 546

\bibitem[{{Mason} {et~al.}(2001){Mason}, {Wycoff}, {Hartkopf}, {Douglass}, \&
  {Worley}}]{2001AJ....122.3466M}
{Mason}, B.~D., {Wycoff}, G.~L., {Hartkopf}, W.~I., {Douglass}, G.~G., \&
  {Worley}, C.~E. 2001, \aj, 122, 3466

\bibitem[{{Menten} {et~al.}(2007){Menten}, {Reid}, {Forbrich}, \&
  {Brunthaler}}]{2007A&A...474..515M}
{Menten}, K.~M., {Reid}, M.~J., {Forbrich}, J., \& {Brunthaler}, A. 2007, \aap,
  474, 515

\bibitem[{{Muterspaugh} {et~al.}(2010){Muterspaugh}, {Lane}, {Kulkarni},
  {Konacki}, {Burke}, {Colavita}, {Shao}, {Hartkopf}, {Boss}, \&
  {Williamson}}]{2010AJ....140.1657M}
{Muterspaugh}, M.~W., {Lane}, B.~F., {Kulkarni}, S.~R., {et~al.} 2010, \aj,
  140, 1657

\bibitem[{{Pallavicini} {et~al.}(1985){Pallavicini}, {Willson}, \&
  {Lang}}]{1985A&A...149...95P}
{Pallavicini}, R., {Willson}, R.~F., \& {Lang}, K.~R. 1985, \aap, 149, 95

\bibitem[{{Sandstrom} {et~al.}(2007){Sandstrom}, {Peek}, {Bower}, {Bolatto}, \&
  {Plambeck}}]{2007ApJ...667.1161S}
{Sandstrom}, K.~M., {Peek}, J.~E.~G., {Bower}, G.~C., {Bolatto}, A.~D., \&
  {Plambeck}, R.~L. 2007, \apj, 667, 1161

\bibitem[{{Torres} {et~al.}(2009){Torres}, {Loinard}, {Mioduszewski}, \&
  {Rodr{\'{\i}}guez}}]{2009ApJ...698..242T}
{Torres}, R.~M., {Loinard}, L., {Mioduszewski}, A.~J., \& {Rodr{\'{\i}}guez},
  L.~F. 2009, \apj, 698, 242

\bibitem[{{van Leeuwen}(2007)}]{2007A&A...474..653V}
{van Leeuwen}, F. 2007, \aap, 474, 653

\bibitem[{{Xu} {et~al.}(2011){Xu}, {Moscadelli}, {Reid}, {Menten}, {Zhang},
  {Zheng}, \& {Brunthaler}}]{2011ApJ...733...25X}
{Xu}, Y., {Moscadelli}, L., {Reid}, M.~J., {et~al.} 2011, \apj, 733, 25

\bibitem[{{Xu} {et~al.}(2006){Xu}, {Reid}, {Zheng}, \&
  {Menten}}]{2006Sci...311...54X}
{Xu}, Y., {Reid}, M.~J., {Zheng}, X.~W., \& {Menten}, K.~M. 2006, Science, 311,
  54

\end{thebibliography}

\begin{deluxetable}{rrrrrr}
\tablecaption{Positions for Secondary Calibrator J2334+2010 \label{tab:secondary}}
\tablehead{
\colhead{JD} & \colhead{Flux} & \colhead{$\alpha$} & \colhead{$\Delta\alpha$} & \colhead{$\delta$} & \colhead{$\Delta\delta$} \\
             & \colhead{(mJy)} & \colhead{(J2000)} & \colhead{(mas)} & \colhead{(J2000)} & \colhead{(mas)} \\
}
\startdata
        2453820 & $  62.2 \pm   1.9 $ & 23 34 14.1564960 & 0.135 & 20 10 28.882640 & 0.279 \\ 
\hline 
        2454482 & $  24.7 \pm   0.2 $ & 23 34 14.1564897 & 0.007 & 20 10 28.882867 & 0.013 \\ 
        2454585 & $  19.6 \pm   0.3 $ & 23 34 14.1564998 & 0.013 & 20 10 28.882864 & 0.022 \\ 
        2454680 & $  13.1 \pm   0.3 $ & 23 34 14.1564977 & 0.018 & 20 10 28.882824 & 0.033 \\ 
        2454793 & $  15.9 \pm   0.3 $ & 23 34 14.1565001 & 0.015 & 20 10 28.882793 & 0.025 \\ 
        2454978 & $  21.4 \pm   0.3 $ & 23 34 14.1564972 & 0.010 & 20 10 28.882838 & 0.017 \\ 
        2455101 & $  23.6 \pm   0.2 $ & 23 34 14.1564926 & 0.007 & 20 10 28.883053 & 0.015 \\ 
        2455226 & $  19.0 \pm   0.1 $ & 23 34 14.1565001 & 0.004 & 20 10 28.882733 & 0.006 \\ 
        2455467 & $  23.8 \pm   0.2 $ & 23 34 14.1565155 & 0.006 & 20 10 28.882853 & 0.010 \\ 
        2455591 & $  21.8 \pm   0.2 $ & 23 34 14.1565109 & 0.006 & 20 10 28.882888 & 0.010 \\ 
\enddata
\tablecomments{The first line is the mean position reported in Paper I.  Errors in the position are errors
from fitting and do not include systematic effects.}
\end{deluxetable}

\begin{deluxetable}{rrrrrr}
\tablecaption{Positions for GJ 896A \label{tab:data}}
\tablehead{
\colhead{JD} & \colhead{Flux} & \colhead{$\alpha$} & \colhead{$\Delta\alpha$} & \colhead{$\delta$} & \colhead{$\Delta\delta$} \\
             & \colhead{($\mu$Jy)} & \colhead{(J2000)} & \colhead{(mas)} & \colhead{(J2000)} & \colhead{(mas)} \\
}
\startdata
\hline 
    2453818.310 & $    929 \pm    203 $ & 23 31 52.4336850 & 0.113 & 19 56 13.712351 & 0.171 \\ 
    2453820.310 & $   1399 \pm    298 $ & 23 31 52.4342630 & 0.183 & 19 56 13.714736 & 0.202 \\ 
    2453821.310 & $   3895 \pm    230 $ & 23 31 52.4345400 & 0.028 & 19 56 13.715419 & 0.045 \\ 
\hline
    2454482.443 & $    328 \pm     20 $ & 23 31 52.4967356 & 0.031 & 19 56 13.570094 & 0.081 \\ 
    2454679.922 & $    278 \pm     25 $ & 23 31 52.5343729 & 0.062 & 19 56 13.704882 & 0.128 \\ 
    2454792.651 & $   2041 \pm     52 $ & 23 31 52.5301598 & 0.015 & 19 56 13.570517 & 0.031 \\ 
    2454978.109 & $   9512 \pm    470 $ & 23 31 52.5703944 & 0.033 & 19 56 13.606525 & 0.060 \\ 
    2455100.776 & $   1139 \pm     33 $ & 23 31 52.5719845 & 0.018 & 19 56 13.599548 & 0.051 \\ 
    2455226.443 & $    138 \pm     16 $ & 23 31 52.5809398 & 0.059 & 19 56 13.443975 & 0.101 \\ 
    2455466.776 & $   2484 \pm     88 $ & 23 31 52.6126099 & 0.021 & 19 56 13.537846 & 0.045 \\ 
    2455591.443 & $   2018 \pm     78 $ & 23 31 52.6215506 & 0.018 & 19 56 13.383628 & 0.044 \\ 
\enddata
\tablecomments{The first three measurements are the results from Paper I.}
\end{deluxetable}

%\begin{deluxetable}{lrrrrrrrrrrr}
%\rotate
%\tabletypesize{\scriptsize}
%\tablecaption{Astrometric Parameters for GJ 896A \label{tab:astro}}
%\tablehead{
%\colhead{Origin} & \colhead{$\alpha$} & \colhead{$\Delta\alpha$} & \colhead{$\delta$} & \colhead{$\Delta\delta$} & \colhead{$\mu_\alpha$} & \colhead{$\mu_\delta$} & \colhead{$\pi$} & \colhead{$a_\alpha$} & \colhead{$a_\delta$} & \colhead{$t_0^\alpha$} & \colhead{$t_0^\delta$} \\
%                 &                   &  \colhead{(mas)} &                  & \colhead{(mas)} & \colhead{(mas/y)}      & \colhead{(mas/y)}      & \colhead{(mas)}  & \colhead{(mas/y$^2$)} & \colhead{(mas/y$^2$)} 
%}
%\startdata
%Optical & 23 31 52.560 & 153 & 19 56 13.90 & 108 & $602.6 \pm 14.0$ & $17.3 \pm 9.7$ & 160 & \dots & \dots& \dots & \dots  \\
%RIPL Astro & 23 31 52.179555 & 0.409 & 19 56 14.141024 & 0.434 & $  571.333 \pm    0.045$  & $  -60.781 \pm    0.047$  & $  159.864 \pm    0.616$ & \dots & \dots  \\ 
%RIPL Astro + Accel & 23 31 52.179565 & 0.123 & 19 56 14.139842 & 0.136 & $  571.169 \pm    0.014$ & $  -60.676 \pm    0.014$ & $  160.010 \pm    0.182$ &  $    0.458 \pm    0.032$ & $    0.087 \pm    0.032$ & $  2454565 \pm       32$ & $  2454879 \pm      167$ \\ 
%\enddata
%\tablecomments{All positions given in ICRS.}
%\end{deluxetable}

\begin{deluxetable}{lrrrr}
%\rotate
%\tabletypesize{\scriptsize}
\tablecaption{Astrometric Parameters for GJ 896A \label{tab:astro}}
\tablehead{
\colhead{Parameter} & \colhead{Unit} & \colhead{Optical} & \colhead{RIPL Astrometric} & \colhead{RIPL Astrometric + Accel} 
}
\startdata
$\alpha_0$ &                &23 31 52.17898       & 23 31 52.179555 &23 31 52.179565 \\
$\Delta\alpha_0$ & (mas)    &12.31                & 0.409 & 0.123 \\
$\delta_0$ &                & 19 56 14.1505       & 19 56 14.141024 & 19 56 14.139842 \\
$\Delta\delta_0$ & (mas)    & 9.51               & 0.434 & 0.136 \\
$\mu_\alpha$ & (mas/y)    &  $554.64 \pm 1.40$ & $  571.333 \pm    0.045$  & $  571.169 \pm    0.014$ \\
$\mu_\delta$ &  (mas/y)   & $-60.43 \pm 1.08$    & $  -60.781 \pm    0.047$  & $  -60.676 \pm    0.014$ \\
$\pi$ &  (mas)            & $ 161.76 \pm 1.66$   & $  159.864 \pm    0.616$ & $  160.010 \pm    0.182$ \\
$a_\alpha$ &  (mas/y$^2$) & \dots             & \dots               & $    0.458 \pm    0.032$ \\
$a_\delta$ &  (mas/y$^2$) & \dots             & \dots               & $    0.087 \pm    0.032$ \\
$t_0^\alpha$ &  (JD)      & \dots             & \dots               & $  2454565 \pm       32$ \\
$t_0^\delta$ & (JD)       & \dots             & \dots               & $  2454879 \pm      167$ \\ 
\enddata
%\tablecomments{All positions given in ICRS.}
\end{deluxetable}

\begin{deluxetable}{rrrrrrr}
\tablecaption{Orbital Parameters for the GJ 896 AB System From Optical Astrometry \label{tab:binary}}
\tablehead{
\colhead{Period} & \colhead{Semi-Major Axis} & \colhead{Inclination} & \colhead{$\Omega$} & \colhead{Periastron} & \colhead{Eccentricity} & \colhead{$\omega$} \\
\colhead{(y)} & \colhead{(arcsec)} & \colhead{(deg)} &\colhead{(deg)} & \colhead{(y)} & & \colhead{(deg)}
}
\startdata
359 &6.87 & 123.5 & 82.1 & 2008 & 0.20 & 354.0 \\
\enddata
%\tablecomments{Source:  Sixth Catalog of Orbits of Visual Binary Stars, Hartkopf \& Mason}
\end{deluxetable}

\end{document}